\documentclass[twocolumn,pre]{revtex4-2}

\usepackage{hyperref}
\usepackage{amsmath}
\usepackage{amsbsy}
\usepackage{amssymb}
\usepackage{graphicx}
\usepackage{xcolor}
\usepackage{bbm}
\usepackage{makecell}
\usepackage{enumerate}
\usepackage{mathtools}

\usepackage{algorithmic}
\usepackage[ruled,vlined,linesnumbered]{algorithm2e}

\newcommand{\sign}{\operatorname{sign}}
\DeclareMathOperator*{\mean}{mean}
\DeclareMathOperator*{\argmin}{arg\,min}

\begin{document}

\title{Graph neural networks and the energetic cavity method \\ for combinatorial optimization}

\author{Joe Bacchus George}
\email{jbg39@cam.ac.uk}
\affiliation{Department of Engineering, University of Cambridge, CB2 1PZ, United Kingdom}

\author{George T. Cantwell}
\email{gtc31@cam.ac.uk}
\affiliation{Department of Engineering, University of Cambridge, CB2 1PZ, United Kingdom}

\begin{abstract}
We study the use of graph neural networks (GNNs) for finding approximate ground states of Ising models. Efficiently finding these ground states is of broad significance because many combinatorial optimization problems can be formulated as an Ising model with the appropriate choice of couplings and fields. Exactly solving these problems is hard but there are many good heuristic methods. A lineage of these heuristics build from mean-field approximations: one approach uses the leading eigenvector of an appropriately defined matrix, another is the min-sum algorithm, also known as the energetic cavity method. Without modification, GNNs perform worse than both of these methods. We consider small modifications to the GNN to incorporate these heuristics and find that this considerably improves performance. While the modified approach is much stronger than other deep learning approaches, we still find that simulated annealing is generally as good for the same computational cost.
\end{abstract}

\maketitle

\section{Introduction}

Combinatorial optimization problems seek a best solution from a finite set of options.
They occur across virtually all areas of science and industry \cite{hartmann_phase_2006}.
Perhaps surprisingly, a finite domain does not make these problems any easier, so the development of practical algorithms for these problems is an important and active field of research.
Often, combinatorial optimization can be recast as finding the ground state of an appropriately defined Ising model \cite{hartmann_phase_2006, moore_nature_2019, barahona_computational_1982}.

It is unlikely that any efficient and general purpose algorithm can truly solve combinatorial problems at least as rich as the Ising model. However, efficient heuristics that find good (but not optimal) solutions certainly do exist and often provide powerful approximations to these problems, e.g., Refs.~\cite{fiedler_algebraic_1973, kirkpatrick_optimization_1983, montanari_solving_2007}.
Heuristic methods build on the intuitions of researchers and often on physical analogies.
Statistical physics has proved particularly relevant for optimization \cite{hartmann_phase_2006}.

Neural networks have been applied to combinatorial optimization tasks \cite{khalil_learning_2017, karalias_erdos_2020, cappart_combinatorial_2023, schuetz_combinatorial_2022, tonshoff_one_2022, zhang_solving_2023}. The efficacy of these approaches remains unclear: neural network methods may find worse solutions than very simple conventional methods and often require considerably more computing resources to achieve this \cite{boettcher_inability_2022, angelini_modern_2022, nath_benchmark_2024}.

Nothing fundamental places conventional methods in opposition to neural networks or deep learning. Indeed, to improve the latter may require thinking carefully about how to incorporate the strengths of the former. For example, the development of convolutional neural networks follows this pattern---the approach built on the understanding that images should be analyzed with local filters, but replaced pre-specified filters with more flexible neural networks \cite{fukushima_neocognitron_1980}.

In this work we incorporate graph neural networks into the min-sum algorithm \cite{mezard_mean-field_1987, mezard_cavity_2003, mezard_information_2009}.
The min-sum algorithm---also known as the energetic cavity method---is an effective heuristic for combinatorial optimization. We construct a neural network so that before training it implements the min-sum algorithm. Starting from this initialization the training procedure should only improve performance.

There is considerable precedent for this approach.
Most relevant are prior studies that incorporated neural networks into the belief propagation (BP) algorithm, which is very closely related to min-sum:
where min-sum attempts to find minima, belief propagation attempts to find marginals in a probabilistic model.
(In the statistical physics setting one can think of the min-sum algorithm as the zero-temperature limit of the belief propagation algorithm, where all probability mass moves to the ground states.)
Several prior studies have considered hybrid neural-network and belief propagation approaches.

At one end of the spectrum, belief propagation is a structural motivation but the equations are entirely replaced by neural networks \cite{yoon_inference_2018, zhang_factor_2020}.
At the other end, belief propagation equations are largely unchanged but a neural network dictates how the between-round updates are computed \cite{kuck_belief_2020}.
Somewhere in the middle, and most similar to our own approach, the belief propagation equations are retained but augmented with a graph neural network operating on the factor graph \cite{satorras_neural_2021}. We follow a similar philosophy, although we build from the min-sum algorithm for optimization and we study problems larger by orders of magnitude.

We draw three conclusions from this work.
First, incorporating well-understood theory into the neural network architecture considerably improves their performance.
Second, despite this improvement, Monte Carlo methods are reliably as good for large problems.
While Monte Carlo algorithms have a reputation for scaling poorly, we find that even fairly aggressive limits on their runtime, e.g., $\mathcal{O}(n \log n)$, scale to better solutions than neural networks.
Finally, our results suggest there is room for an improved physics-of-ground-state-energies, i.e., there exists a set of equations that improve on the energetic cavity equations.
For now these are obscure---in the form of neural networks---but in time we hope to understand this better.

\section{Problem and model setup}
We consider optimization problems of the form
\begin{equation}
	\argmin_{\boldsymbol{\sigma} \in \{-1, +1\}^{n}} \Big\{ -\frac{1}{2}\sum_{i=1}^{n}\sum_{j=1}^{n} J_{ij}\sigma_i \sigma_j - \sum_{i=1}^n h_i \sigma_i  \Big\},
\end{equation}
or equivalently, finding ground states of an Ising model with Hamiltonian
\begin{equation}
	H(\boldsymbol{\sigma}) = - \frac{1}{2}\sum_{i,j} J_{ij} \sigma_i \sigma_j - \sum_i h_i \sigma_i.
\label{eq:Ising_hamiltonian}
\end{equation}
This class of problems is sufficiently rich to be of broad interest and is arguably the simplest class of non-trivial optimization problems on graphs; it is also NP-hard \cite{Karp1972, barahona_computational_1982}.

A majority of our experiments consider the MaxCut problem. Namely, given a graph, we attempt to sort the nodes into two groups so that the largest possible number of edges fall between nodes in different groups. Finding the maximum cut of graph $G$ is equivalent to minimizing
\begin{equation}
	\sum_{i<j} A_{ij} \sigma_i \sigma_j
\label{eq:maxcut}
\end{equation}
where $A$ is the adjacency matrix so that the sum is equal to the number of edges within groups minus the number of edges between the two groups. The MaxCut problem corresponds to an Ising model where all $h_i=0$ and $J_{ij} = -A_{ij}$, i.e., an anti-ferromagnet in the absence of an external field.

\subsection{Graph neural networks}
Minimizing Eq.~\eqref{eq:Ising_hamiltonian} is difficult. 
At the same time, neural networks have proven themselves capable for many non-trivial problems. Graph neural networks (GNNs) are the leading paradigm for applying neural networks to graphs \cite{scarselli_graph_2009, kipf_semi-supervised_2016, hamilton_inductive_2017} and, unsurprisingly, attempts have been made to use GNNs to solve Ising-like combinatorial problems \cite{khalil_learning_2017, karalias_erdos_2020, cappart_combinatorial_2023, schuetz_combinatorial_2022, tonshoff_one_2022}.

Graph neural networks generate hidden embeddings for nodes in a graph. A $T$ layer GNN of width $d$ computes embeddings $\boldsymbol{y}_i^{(1)}, \dots, \boldsymbol{y}_i^{(T)} \! \in \mathbb{R}^{d}$ for each node $i$, starting with an initial condition, e.g., $\boldsymbol{y}_i^{(0)}=\left[1.0\right]$, and  performing updates according to
\begin{equation}
\boldsymbol{y}_i^{(t+1)} = f \Big( \overline{\boldsymbol{b}}^{(t)} + \overline{W}_1^{(t)} \boldsymbol{y}_i^{(t)} + \overline{W}_2^{(t)} \bigoplus\nolimits_{j} J_{i j} \boldsymbol{y}_j^{(t)} \Big)
\end{equation}
where $f$ is an element-wise non-linear function, $\bigoplus_{j}$ is an aggregation over nodes, and the quantities with overlines ($\overline{\boldsymbol{b}}^{(t)}$, $\overline{{W}}_1^{(t)}$, $\overline{{W}}_2^{(t)}$) are the free parameters that are ``learned''.

To apply this to the Ising problem of Eq.~\eqref{eq:Ising_hamiltonian} we set $\boldsymbol{y}_i^{(0)}$ to be normally distributed random noise and then sequentially perform updates
\begin{equation}
	\boldsymbol{y}_i^{(t+1)} = \tanh \Big( \overline{\boldsymbol{b}}^{(t)} + \overline{W}_1^{(t)} \boldsymbol{y}_i^{(t)} + \overline{W}_2^{(t)}  \mean_{j \in \partial_i }\big\{ J_{i j} \boldsymbol{y}_j^{(t)} \big\} \Big)
\label{eq:gnn}
\end{equation}
where $\partial_i$ is the set of nodes adjacent to $i$.
An assignment to spins is extracted at the final layer by
\begin{equation}
	\sigma_i = \sign \Big( \overline{\boldsymbol{w}}_{\mathrm{(out)}} \cdot \boldsymbol{y}_i^{(T)} \Big)
\end{equation}
where again, the overline on $\overline{\boldsymbol{w}}_{\mathrm{(out)}}$ indicates a vector of free parameters.

For a GNN to be useful we must be able to train the weights, i.e., fit the free parameters. Standard methods to achieve this require a differentiable loss function. To maintain differentiability we replace $\sigma_i$, which are binary and therefore non-differentiable, with the relaxation
\begin{equation}
	z_i = \tanh \Big( \overline{\boldsymbol{w}}_{\mathrm{(out)}} \cdot \boldsymbol{y}_i^{(T)} \Big).
\end{equation}
We can then insert these relaxed values $z_i$ into our energy function, $H \! \left(\boldsymbol{z}\right)$, and differentiate with respect to the neural network weights. 
Different classes of problem correspond to different distributions of $J$ and $\boldsymbol{h}$.
Once these distributions are specified we attempt to minimize
\begin{equation}
\mathcal{L}(\boldsymbol{\theta}) = \mathbb{E}_{J, \boldsymbol{h}} \! \left[ H \! \left(\boldsymbol{z}_{\boldsymbol{\theta}} \! \left(J, \boldsymbol{h} \right)\right) \right],
\label{eq:full_objective}
\end{equation}
where $\boldsymbol{z}_{\boldsymbol{\theta}} \! \left(J, \boldsymbol{h}\right)$ is the output of the GNN with parameters ${\boldsymbol{\theta}}$ on problem $J, \boldsymbol{h}$.
For example, when we consider MaxCut on a random $3$-regular graph, we set $\boldsymbol{h} = \boldsymbol{0}$ and $J = -A$, where $A$ is the adjacency matrix of a random $3$-regular graph.

The expectation in Eq.~\eqref{eq:full_objective} is intractable but we can still minimize it using stochastic methods. To approximate the gradient we sample a set of $N$ problems
$\{ (J_\alpha, \boldsymbol{h}_\alpha) \}_{\alpha=1}^N$
from the appropriate distribution and then approximate the gradient by 
\begin{equation}
\frac{\mathrm{d}\mathcal{L}}{\mathrm{d}\boldsymbol{\theta}}
\simeq
\frac{1}{N}\sum_{\alpha=1}^{N} \frac{\mathrm{d} H \! \left(\boldsymbol{z}_{\boldsymbol{\theta}} \! \left(J_\alpha, \boldsymbol{h}_\alpha \right) \right)}
{\mathrm{d} \boldsymbol{\theta}}.
\end{equation}
Across all experiments, we use the training procedure of Appendix~\ref{app:training}. The approach is unsupervised, i.e., does not rely on solved instances for training.
This is important precisely because the optimization is difficult: we can easily generate new problems but we cannot easily generate their solutions. If one could already generate a large enough set of solutions to train a neural network then, presumably, one does not need the neural network to solve the problem in the first place.

\section{GNN experiments}
To explore the suitability of GNNs for optimization let us initially restrict ourselves to MaxCut on random graphs. Maximum cut for the random $k$-regular graph has been carefully studied, both numerically and theoretically, and so this provides a useful testing ground for GNNs \cite{mezard_cavity_2003, boettcher_numerical_2003, zdeborova_conjecture_2010, dembo_extremal_2017}.

Prior studies found mixed to poor results when GNNs are used for optimization.
Some architectures even appear to be worse than the most basic greedy algorithms \cite{boettcher_inability_2022, angelini_modern_2022, nath_benchmark_2024}. Because of this,
and ahead of reporting any GNN-based results, we first consider an appropriate baseline for comparison. 

\subsection{Laplacian cut as a baseline}
A simple heuristic for MaxCut uses the graph Laplacian \cite{fiedler_algebraic_1973, trevisan_max_2012}. Intuitively, the eigenvectors of the Laplacian with the largest eigenvalues correspond to signals that are oscillating most rapidly.
Define the normalized Laplacian as
\begin{equation}
  L =  I - D^{-1} A
\end{equation}
where $I$ is the identity and $D$ is the diagonal matrix of node degrees.
Then, let $\boldsymbol{v}^{\mathrm{max}}(L)$ be an eigenvector corresponding to the largest eigenvalue of $L$.
The Laplacian heuristic finds a 
cut by assigning
\begin{equation}
  \sigma_i = \sign \left( {v}^{\mathrm{max}}_i(L) \right).
  \label{eq:lapl_cut}
\end{equation}
This method is simple to implement and is free of parameters or further optimizations. 
At the very least, any method of practical interest should consistently outperform this elementary heuristic.

To find the leading eigenvector of $L$ one can apply the power iteration method.
Namely, begin by initializing a vector $\boldsymbol{x}^{(0)}$ at random. Then, repeatedly multiply by the Laplacian to get $\boldsymbol{x}^{(t+1)} = L \boldsymbol{x}^{(t)}$. After a sufficiently large number of iterations $T$ the vector $\boldsymbol{x}^{(T)}$ will be very close to the largest eigenvector of $L$. Writing out the matrix multiplication explicitly we have
\begin{equation}
x_i^{(t+1)} = x_i^{(t)} - \mean_{j \in \partial_i} \left\{ x_j^{(t)} \right\}
\end{equation}
and comparing to Eq.~\eqref{eq:gnn} one sees a clear similarity.
In fact, if we add an element-wise non-linearity we recover the mean-field approximation
\begin{equation}
{x}_i^{(t+1)} = \tanh\Big( {x}_i^{(t)} -  \mean_{j \in \partial_i} \left\{ {x}_j^{(t)} \right\} \Big)
\label{eq:lapl_nonlinear}
\end{equation}
which is precisely a GNN of width $d=1$ and parameters
\begin{equation}
\overline{\boldsymbol{b}}^{(t)} = 0; \quad \overline{W}^{(t)}_1 = -\overline{W}^{(t)}_2= 1; \quad \overline{\boldsymbol{w}}_3 = 1.
\end{equation}
Hence, the Laplacian heuristic essentially is a GNN of unit width and no free parameters. It is consequently trivial that there \emph{exists} a GNN which performs at least as well as the Laplacian on maximum cut. The question then becomes whether or not the GNN can make any improvement on this baseline.

\subsection{GNN depth and weight initialization}

Graph neural networks update node representations from $\boldsymbol{y}_i^{(0)}$ to 
$\boldsymbol{y}_i^{(T)}$ through $T$ layers. 
Folk wisdom advocates that $T$ should be relatively small and typical GNN architectures tend to be no deeper than $8$ layers. The justification is that, in practice, training seems to fail for deep GNNs as node representations suffer from ``oversmoothing''. That is, information repeatedly diffuses between nodes so that their embeddings become almost identical \cite{li_deeper_2018, rusch_survey_2023}.

A shallow GNN only allows nodes to share information with others that are nearby.
While shallow GNNs might be the preferred method of practice, the physics of a problem such as MaxCut pushes against this. In sparse graphs we typically expect the diameter to grow logarithmically in $n$ \cite{newman_networks_2018} and local methods are unlikely to find optimal solutions \cite{chen_suboptimality_2019, gamarnik_barriers_2023}. For information to propagate across macroscopic distances in sparse graphs, the physical intuition remains that we need deep rather than shallow networks, presumably growing in depth at least as fast as $\log(n)$.

\begin{figure*}
  \centering
  \includegraphics[width=7in]{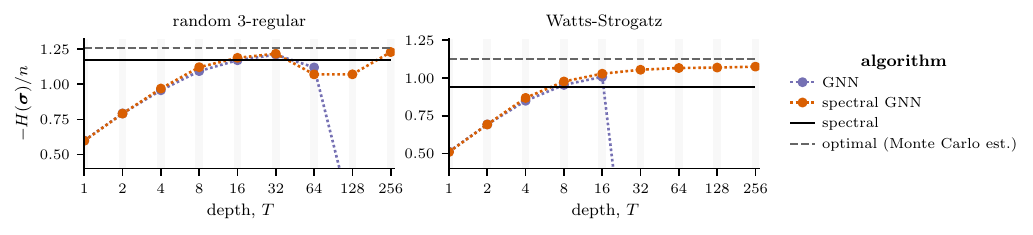} 
  \caption{Performance of GNNs with different numbers of layers: their depth.
	We train GNNs to solve MaxCut on random $3$-regular graphs and on Watts-Strogatz graphs ($k=4$ and $p=1/10$) of size $8192$. 
	Results are obtained by computing the average solution on $100$ graph samples of size $n=8192$ for each depth considered, taking the best solution out of $10$ attempts for every method.
	We show performance of two GNN methods, one initialized with standard weights (GNN) and one initialized so that it performs a Laplacian cut before training (spectral GNN).
	With an informed choice for initialization, deep models with $256$ layers are able to considerably outperform a standard Laplacian spectral cut. 
	}
  \label{fig:plot_0}
\end{figure*}

Our first numerical experiment, in Fig.~\ref{fig:plot_0}, shows the results of training a standard GNN for MaxCut on random $3$-regular graphs, and on Watts-Strogatz graphs \cite{watts_collective_1998}, with $n=8192$ nodes for both cases.

At shallow depths---typical for standard GNN architectures---the GNN is unable to beat the Laplacian baseline. At least for this task the shallow GNN framework appears to be worse than useless: it introduces considerable computation cost, and for that price it considerably reduces performance. This is concordant with the findings of Refs.~\cite{schuetz_combinatorial_2022, boettcher_inability_2022, angelini_modern_2022}.

As the depth is increased, GNN performance initially improves and by $16$ layers the GNN starts to appear competitive against the trivial baseline.
Unfortunately, performance then falls off a cliff. For truly deep architectures, which the physics of the problem necessitates if we want strong performance on very large graphs, the GNN performance collapses back to chance.

However, the failure of the deep GNN to perform at least as well as the Laplacian must be due to a poor choice of weights---it is a failure of learning.
This must be true because the Laplacian method is equivalent to a very deep GNN with no free parameters.

The Laplacian argument shows that we already know a set of weights that perform well, even before training.
Defining $\mathbf{e}_1$ to be the first standard basis vector and $P = I - \mathbf{e}_1 \mathbf{e}_1^\top$, we reinitialize the GNN weights as
\begin{equation}
\begin{aligned}
\overline{\boldsymbol{b}}^{(t)} &\leftarrow P \overline{\boldsymbol{b}}^{(t)}  \\
\overline{W}^{(t)}_1 &\leftarrow P \overline{W}^{(t)}_1 P + \mathbf{e}_1 \mathbf{e}_1^\top \\
\overline{W}^{(t)}_2 &\leftarrow P \overline{W}^{(t)}_2 P - \mathbf{e}_1 \mathbf{e}_1^\top \\
\overline{\boldsymbol{w}}_3 &\leftarrow  \mathbf{e}_1
\end{aligned}
\label{eq:lapl_init}
\end{equation}
and the effect of these initializations is that, before training, the GNN is equivalent to the iterative implementation of the Laplacian cut in Eq.~\eqref{eq:lapl_nonlinear}. That is to say, we begin with a model whose performance is at least no worse than a reasonable baseline.

The question remains as to whether this GNN can learn anything beyond the Laplacian heuristic we give it at initialization. Our experiments answer this in the affirmative: when the GNN is initialized so that it performs Laplacian MaxCut it is capable of considerable further improvement. In Fig.~\ref{fig:plot_0} we see that the GNN initialized by Eq.~\eqref{eq:lapl_init}, denoted as spectral GNN, improves over the baseline method at depth $256$.
To estimate the quality of a method we average over $100$ problem instances and on each instance we use the best of $10$ trials.

So far no changes have been proposed to the neural architecture---only the weight initialization was modified for the spectral GNN. The fact that these deep networks improve on the original method is noteworthy, namely, it emphasizes the importance of incorporating heuristic knowledge into the GNN model design. In the next section we consider a more significant change to the architecture.
Specifically, we attempt to improve the min-sum algorithm which is a considerably more competitive method than the Laplacian approach on locally tree-like graphs.

\section{Min-Sum experiments}

The min-sum algorithm, also known as the energetic cavity equations \cite{mezard_cavity_2003, mezard_information_2009}, is well-established for finding approximate ground states in Ising-like models.


To derive the equations of the min-sum algorithm, denote by $E_i(\pm)$ the minimum energy of Hamiltonian~\eqref{eq:Ising_hamiltonian} when $\sigma_i=\pm 1$,
\begin{align}
E_i(+) &= \min_{\boldsymbol{\sigma} : \sigma_i = + 1}H(\boldsymbol{\sigma}), \label{eq:E_i_plus} \\
E_i(-) &= \min_{\boldsymbol{\sigma} : \sigma_i = - 1}H(\boldsymbol{\sigma}). \label{eq:E_i_minus}
\end{align}
The optimal assignment of $\sigma_i$ in fact only depends on the difference between these, and so defining
\begin{equation}
  \nu_i = \frac{1}{2} \big(E_i(-) - E_i(+)\big)
\end{equation}
the optimal choice is to set
\begin{equation}
  \sigma_i = \sign(\nu_i).
\end{equation}
The problem, of course, is that we don't know $\nu_i$.

To make further progress let us consider a modified problem in which node~$i$ is removed.
In this modified problem, define $E_{i \leftarrow j}(\pm)$ to be the minimum energy achievable when $\sigma_j = \pm 1$. 
If the graph were a tree (i.e., contained no cycles) we could write
\begin{align}
E_i(+) &= -h_i + \sum_{j \in \partial_i} \min \big(E_{i \leftarrow j}(+)-J_{ij}, E_{i \leftarrow j}(-) + J_{ij}\big) \nonumber \\
E_i(-) &= +h_i + \sum_{j \in \partial_i} \min \big(E_{i \leftarrow j}(+) + J_{ij}, E_{i \leftarrow j}(-) -J_{ij}  \big)
\end{align}
and so defining the function
\begin{equation}
  \phi(x, y) = \sign( x y) \min( \vert x \vert, \vert y \vert)
\end{equation}
we get
\begin{align}
\nu_i = h_i + \sum_{j \in \partial_i} \phi(J_{ij}, \nu_{i \leftarrow j} )
\label{eq:marginal_nu}
\end{align}
where $\nu_{i \leftarrow j} = (E_{i \leftarrow j}(-) - E_{i \leftarrow j}(+))/2$.

At a first glance it might not be clear why Eq.~\eqref{eq:marginal_nu} is helpful because we again do not know the values $\nu_{i \leftarrow j}$. The insight is that, on a tree, these can likewise be written as
\begin{equation}
\nu_{i \leftarrow j} = h_j + \sum_{k \in \partial_j \setminus i} \phi(J_{jk}, \nu_{j \leftarrow k} ).
\end{equation}
Combining these with Eq.~\eqref{eq:marginal_nu}, we have the following set of closed equations
\begin{equation}
\begin{aligned}
	\nu_i &= h_i + \sum_{j \in \partial_i} \mu_{i \leftarrow j} \\
	\mu_{i \leftarrow j} &= \phi(J_{ij}, \nu_j - \mu_{j \leftarrow i})
\end{aligned}
\label{eq:minsum_updates}
\end{equation}
and since these are closed, they can be solved by an iterative method. 
Namely, we can initialize all the variables $\mu_{i \leftarrow j}$ at random, and them iteratively update both $\mu_{i \leftarrow j}$ and $\nu_i$ with Eq.~\eqref{eq:minsum_updates}. 
At convergence, we have a set of equations that give the exact ground state on a tree. This protocol is the min-sum algorithm.

While the equations are only strictly correct on a tree, in practice they can give good solutions on other topologies, particularly locally tree-like graphs. One caveat is that updating by Eq.~\eqref{eq:minsum_updates} can cause oscillations so one typically introduces damping to the equations, which we set to $1/2$ in all of our experiments, as shown in Algorithm~\ref{alg:MS}.

\begin{algorithm}[t]
\caption{min-sum algorithm (\textbf{MS})}
\label{alg:MS}

\KwIn{graph $G=(V,E)$, couplings $J$, fields $\boldsymbol{h}$, iterations $T$}
\KwOut{cavity fields $\boldsymbol{\nu}$}

\textbf{Define} $\phi(a,b) = \sign(a b) \min(|a|,|b|)$

  \medskip

initialize $\mu_{i \leftarrow j}^{(0)} \sim \mathcal{N}(0,1)$, $\forall i \in V, j \in \partial_i$

$\nu_{i}^{(0)} = \sum_{j \in \partial_i} \mu_{i \leftarrow j}^{(0)}$, $\forall i \in V$

\For{$t = 0$ \KwTo $T-1$}{
  \ForEach{$i \in V$}{
	  \ForEach{$j \in \partial_i$}{
		  $ \displaystyle \mu_{i \leftarrow j}^{(t+1)} \leftarrow \frac{1}{2} \Big( \mu_{i \leftarrow j}^{(t)} +  \phi(J_{i j},\nu_{j}^{(t)} - \mu_{j \leftarrow i}^{(t)}) \Big) $
  }
	$ \displaystyle \nu_{i}^{(t+1)} \leftarrow  h_i + \sum_{j \in \partial_i} \mu_{i \leftarrow j}^{(t+1)}$
  }
}

\Return $\boldsymbol{\nu}^{(T)}$
\end{algorithm}

The min-sum algorithm with damping, Algorithm~\ref{alg:MS}, estimates the differences between the lowest energies achieved when $\sigma_i = \pm 1$.
To extract a prediction for the best solution $\boldsymbol{\sigma}$, we take $\sigma_i = \sign(\nu_i)$.

On graphs with cycles, where the min-sum algorithm makes mistakes, it is usually better to extract answers via decimation: sequentially fix a proportion of spins after each round, and then re-run the algorithm with these spins fixed \cite{montanari_solving_2007}. Decimation and the min-sum algorithm generally do not find true minima but are a competitive heuristic. Looking ahead, to Fig.~\ref{fig:plot_2_RR}, we see that the min-sum algorithm considerably outperforms the spectral GNN on random regular graphs; it also outperforms several other deep learning-based methods.

The min-sum algorithm is derived from an incorrect assumption that the graph is a tree. For this reason, on locally tree-like graphs decimated min-sum is expected to find solutions reasonably close to optimal. Whereas, on graphs that are rich with short cycles the procedure is expected to be considerably less good.
How to systematically correct for this shortcoming is an interesting question in its own right \cite{yedidia_constructing_2005, chertkov_loop_2006, cantwell_message_2019}. The particular question we ask now is whether a GNN can build from min-sum to make improvements that correct for its shortcomings.

\subsection{Min-Sum GNN}
The message passing formalism of the min-sum algorithm naturally lends itself to a GNN-like construction.
To this end, we adjust the min-sum messages of Eq.~\eqref{eq:minsum_updates} by including $d$-dimensional features $\boldsymbol{y}_i \in \mathbb{R}^{d}$ at each node
(in all experiments we fix $d=32$).

Written sequentially, the standard min-sum equations are
\begin{equation}
\begin{aligned}
\mu_{i \leftarrow j}^{(t+1)} &= \frac{1}{2} \Big(\mu_{i \leftarrow j}^{(t)} + \phi(J_{ij}, \nu_j^{(t)} - \mu_{j \leftarrow i}^{(t)}) \Big) \\
\nu_i^{(t+1)} &= h_i + \sum_{j \in \partial_i} \mu_{i \leftarrow j}^{(t+1)}
\end{aligned}
\end{equation}
and we initialize $\mu_{i \leftarrow j}^{(0)} \sim  \mathcal{N}(0,1)$.
Whereas, with the addition of a GNN feature which we initialize $\boldsymbol{y}_i^{(0)} = (1, 1, \dots, 1)$, the update equations become
\begin{equation}
\begin{aligned}
	\boldsymbol{y}_i^{(t+1)} &= \tanh \Big( \overline{\boldsymbol{b}}^{(t)} +   \nu_i^{(t)}\overline{\boldsymbol{u}}^{(t)} + \overline{W}_1^{(t)} \boldsymbol{y}_{i}^{(t)} \\ &\quad\quad\quad\quad\quad\quad\quad\quad\quad\quad + \overline{W}_2^{(t)} \mean\limits_{j \in \partial_i} \big\{J_{i j} \boldsymbol{y}_{j}^{(t)} \big\} \Big) \\
	\mu_{i \leftarrow j}^{(t+1)} &= \frac{1}{2} \big(\mu_{i \leftarrow j}^{(t)} + \phi(J_{ij}, \nu_j^{(t)} - \mu_{j \leftarrow i}^{(t)} + \overline{\boldsymbol{w}}^{(t)} \cdot \boldsymbol{y}_j^{(t)}) \big) \\
	\nu_i^{(t+1)} &= h_i + \sum_{j \in \partial_i} \mu_{i \leftarrow j}^{(t+1)}
\end{aligned}
\label{eq:minsum_GNN}
\end{equation}
and as before, overlines indicate neural network parameters which are optimized for the task.

After running through $T$ layers, the final output is a linear combination of $\nu_i^{(T)}$ and $\boldsymbol{y}_i^{(T)}$
\begin{equation}
	\nu_i^{\mathrm{(out)}} = \nu_i^{(T)} + \overline{\boldsymbol{w}}_{\mathrm{(out)}} \cdot \boldsymbol{y}_i^{(T)}
\end{equation}
and spins can be assigned $\sigma_i = \sign(\nu_i^{\mathrm{(out)}})$.
The parameters $\overline{\boldsymbol{w}}^{(t)}$ and $\overline{\boldsymbol{w}}_{\mathrm{(out)}}$ are initialized to zeros so as to pass through the min-sum calculation unaltered, i.e., before training the GNN simply performs the min-sum algorithm with $T$ iterations.
Again, the question at hand is whether the neural networks can improve on min-sum alone.

\subsection{MaxCut on random regular graphs}
\begin{figure}
    \centering
    \includegraphics[width=0.975\columnwidth]{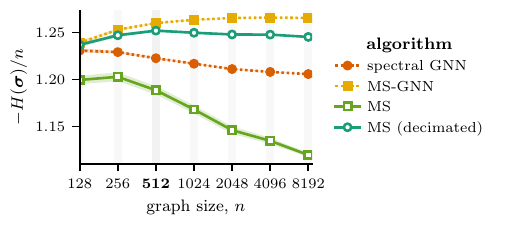}
	\caption{Solution quality for MaxCut on random $3$-regular graphs. The min-sum-GNN performs better at all tested sizes than min-sum alone (with or without decimation). The standard GNN architecture is systematically worse than the min-sum with decimation.
	Both GNN methods were trained at $n=512$; the tests demonstrate that the GNNs are able to extrapolate to unseen sizes.
	}
    \label{fig:plot_2_RR}
\end{figure}

In Fig.~\ref{fig:plot_2_RR} we compare the min-sum-GNN of Eq.~\eqref{eq:minsum_GNN} to the Laplacian GNN model, and to the standard min-sum itself on random $3$-regular graphs. Three things are worthy of note.

First, the decimated min-sum algorithm outperforms the deep GNN methods considered previously. This again illustrates how GNNs are not a panacea and, without adjustment, underperform sensible traditional algorithms.

Second, the min-sum-GNN outperforms all other methods so far considered, including the original min-sum algorithm. We believe that for this task the min-sum-GNN is a genuinely competitive heuristic for performance.

Third, in this figure, the methods are trained on graphs with $n=512$ nodes but appear to generalize well to considerably larger graphs.
The min-sum algorithm is well-suited to the random regular graphs because these graphs have vanishing densities of short cycles, that is, they are locally tree-like. Still, Fig.~\ref{fig:plot_2_RR} demonstrates that without decimation the min-sum algorithm is performing less well on larger graphs. Whereas, with decimation the performance of min-sum appears roughly consistent across different graph sizes.
The min-sum-GNN is built atop the standard min-sum algorithm (without decimation).
Therefore, it is clear from Fig.~\ref{fig:plot_2_RR} that the GNN part of the min-sum-GNN is learning something significant. Not only does it correct for the failure of the non-decimated min-sum approach, but it even outperforms decimation. 

To explore how truly competitive the min-sum-GNN is, we now compare it to simulated annealing---one of the best general purpose methods for optimization problems \cite{kirkpatrick_optimization_1983}. How does this deep learning model actually compete against a veteran heuristic? The question is somewhat poorly posed because one can always increase the run-time of simulated annealing to increase the quality of solutions. 

\begin{figure}
    \centering
    \includegraphics[width=0.975\columnwidth]{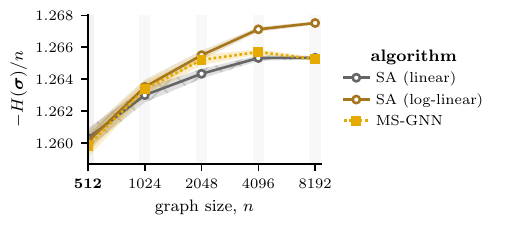}
    \caption{Average solution quality for MaxCut on random $3$-regular graphs with differing numbers of nodes. 
    The min-sum-GNN performs comparably to simulated annealing with a linear run time.
	When simulated annealing is allowed to run in log-linear time, $\mathcal{O}(n \log n)$, it finds better solutions. 
	}
    \label{fig:SA_GNN_comparison}
\end{figure}

In order to conduct a fair comparison between the min-sum-GNN and simulated annealing, we must fix the run time of the latter approach. To do so, we run simulated annealing for $c n$ steps, where $c$ is chosen so that its performance is comparable to the min-sum-GNN when $n=512$. In Fig.~\ref{fig:SA_GNN_comparison}, we show how the performance of the min-sum-GNN and simulated annealing run for $c n$ steps change as we increase the number of nodes, $n$. Interestingly, we find relatively similar scalings in performance. That is to say, min-sum-GNN---which runs in linear time on this class of graphs---performs comparably to simulated annealing with the same resources.

We also test min-sum-GNN against simulated annealing with $\mathcal{O}(n \log n)$ steps. The results in Fig.~\ref{fig:SA_GNN_comparison} show that simulated annealing with runtime growing as $n \log n$ saturates at a considerably higher level than linear time simulated annealing or the min-sum-GNN.

\begin{figure*}
    \centering
    \includegraphics[width=7in]{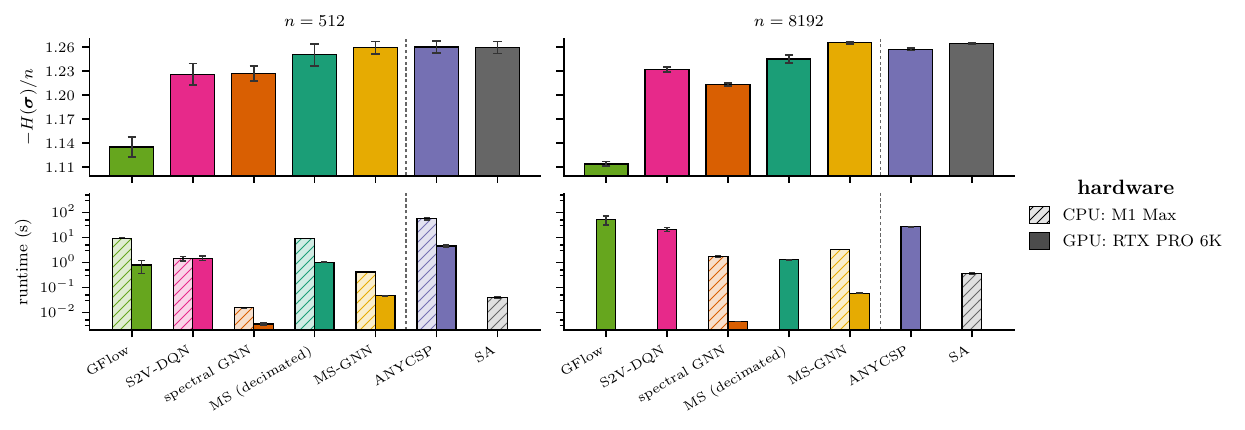}
    \caption{Performance of heuristics on MaxCut for random $3$-regular graphs. We report run times in seconds on a CPU (Macbook Pro M1 Max) and GPU (Nvidia Blackwell RTX Pro 6000). 
    Results are averaged over $100$ problem instances; for the random methods we use the best of $10$ trials on each instance.
    ANYCSP and simulated annealing are search-based methods and both find good solutions, though simulated annealing is much faster.
    All GNN methods except for min-sum-GNN find significantly worse solutions than simulated annealing or even the min-sum algorithm.
    In addition to finding worse solutions, GFlowNets and S2V-DQN are slower than simulated annealing.
    }
    \label{fig:MSGNN_other_methods}
\end{figure*}

Comparing to other deep learning approaches the min-sum-GNN emerges very favorably.
Figure~\ref{fig:MSGNN_other_methods} compares against GFlowNets \cite{zhang_solving_2023}, S2V-DQN  \cite{khalil_learning_2017}, and ANYCSP \cite{tonshoff_one_2022}; our implementations are adapted from Ref.~\cite{nath_benchmark_2024}, which reviewed performance of many different GNN methods.

GFlowNets and S2V-DQN considerably under-perform in solution quality while also taking more than $1000$ times longer to run.
They are simply not competitive.

ANYCSP finds good solutions.
Unlike the other GNN methods ANYCSP takes a search based approach, similar to simulated annealing.
Where simulated annealing proposes small moves at random, ANYCSP learns to propose good moves, i.e., moves that both change a large number of spins and improve the objective considerably.
The solutions found by ANYCSP are competitive with both min-sum-GNN and simulated annealing but these solutions are found more than $500$ times slower than simulated annealing.

All GNNs have additional training expenses and unpredictable performance on out-of-distribution problems.
Simulated annealing has no equivalent issues and remains the most reliable heuristic for this task.

In summary, on random regular graphs, the min-sum-GNN appears competitive against simple heuristics and other learned methods. 
The min-sum-GNN appears to have learned something and corrects for some weaknesses in the min-sum heuristic.
Random regular graphs are locally tree-like and so min-sum style algorithms are expected to do fairly well. If a GNN can in principle correct problems inherent to the min-sum algorithm, can it correct one of the major problems, namely, can it learn to correctly account for the cycles in the graph?

\subsection{MaxCut on Watts-Strogatz graphs}
The Watts-Strogatz model provides a comparable set of graphs to the random regular experiments but, crucially, the graphs are not locally tree-like \cite{watts_collective_1998}.
The model starts with a $1$-dimensional lattice where each node is connected to its $k$ nearest nodes. Then, for each edge $(i,j)$, the endpoint is randomly reassigned from $j$ to a different node, with probability $p$. 
When $p=0$ we obtain a perfectly ordered lattice with lots of short cycles; when $p=1$ we have a locally tree-like graph with a logarithmic diameter. For intermediate re-wiring probabilities, e.g., $p=1/10$, the generated graphs contain lots of triangles while also having logarithmic diameters. The rewiring probability $p$ controls how locally tree-like the networks are---larger $p$ means fewer short cycles.

\begin{figure}
    \centering
    \includegraphics[width=0.975\columnwidth]{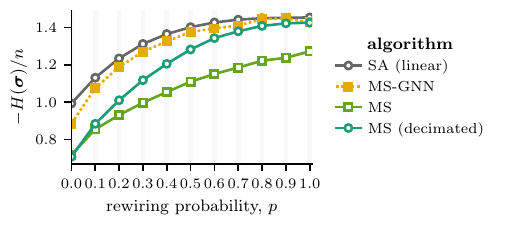}
	\caption{MaxCut solution quality for each algorithm as a function of the rewiring probability, $p$, in the Watts-Strogatz graph with $n=8192$ and $k=4$.
	The min-sum algorithm has weaker performance across a wide range of $p<1$. 
	The min-sum-GNN considerably outperforms traditional min-sum algorithms, but still underperforms simulated annealing.
	}
    \label{fig:WS_MaxCut}
\end{figure}

Figure~\ref{fig:WS_MaxCut} displays the MaxCut results for different methods against the rewiring probability, $p$. Both variants of the min-sum algorithm perform poorly compared to simulated annealing across a wide range of $p$. This is unsurprising because for any $p < 1$ the graphs are not locally tree-like and so the methodological premise of the min-sum algorithm is broken.

In contrast to the standard min-sum algorithms, the min-sum-GNN performs fairly well over the whole range of $p$. At the same time, the min-sum-GNN still noticeably underperforms simulated annealing. The implication is that, while the neural networks are able to improve on the baseline heuristics they incorporate, they are still somewhat constrained by their shortcomings.

\subsection{Further tasks}
We now consider several more tasks and topologies. In addition to the random regular and Watts-Strogatz graphs, we look at random recursive trees and Barabasi-Albert graphs \cite{barabasi_emergence_1999}.
On these topologies we study four different problems:
\begin{enumerate}[(i)]
	\item Maximum cut (MaxCut), where $J_{ij}=-A_{ij}$ and $h_i=0$.
	\item Binary spin glass (BinSG), where each $J_{ij}=\pm A_{ij}$ with the sign chosen independently with probability $1/2$, and $h_i=0$.
	\item Edwards-Anderson (EA), where each $J_{ij} = A_{ij} \varepsilon_{ij}$ with $\varepsilon_{ij}$ drawn independently from a standard normal distribution, and $h_i=0$.
	\item Random field Ising model (RFIM) with $J_{ij} = A_{ij}$ and $h_i$ drawn independently from a standard normal distribution.
\end{enumerate}
Results for all $16$ combinations of problems and topologies are shown in Table~\ref{tab:results}.

\begin{table*}
  \centering
  {\setlength{\tabcolsep}{12pt}
\begin{tabular}{l|ccc|c} \hline
	\makecell[c]{\textbf{Algorithm}} & \makecell[c]{\textcolor[HTML]{666666}{$\bullet$} SA (constrained runtime)} & \makecell[c]{\textcolor[HTML]{66a61e}{$\bullet$} min-sum} & \makecell[c]{\textcolor[HTML]{e6ab02}{$\bullet$} min-sum-GNN} & \makecell[c]{GNN improvement} \\ \hline
\hline \multicolumn{5}{l}{\textbf{MaxCut} } \\ \hline
\quad RT& $0.994 \pm 0.001$ & $1.000 \pm 0.000$ & $1.000 \pm 0.000$ & $*$  \\
\quad RR& $1.263 \pm 0.001$ & $1.123 \pm 0.021$ & $1.267 \pm 0.001$ & $102.5\%$  \\
\quad BA& $1.347 \pm 0.003$ & $1.197 \pm 0.040$ & $1.345 \pm 0.005$ & $98.5\%$  \\
\quad WS& $1.130 \pm 0.003$ & $0.852 \pm 0.006$ & $1.075 \pm 0.004$ & $80.3\%$  \\
\hline \multicolumn{5}{l}{\textbf{BinSG} } \\ \hline
\quad RT& $0.994 \pm 0.001$ & $1.000 \pm 0.000$ & $1.000 \pm 0.000$ & $*$  \\
\quad RR& $1.264 \pm 0.001$ & $1.123 \pm 0.020$ & $1.266 \pm 0.001$ & $101.7\%$  \\
\quad BA& $1.348 \pm 0.003$ & $1.216 \pm 0.044$ & $1.356 \pm 0.004$ & $106.5\%$  \\
\quad WS& $1.372 \pm 0.005$ & $1.143 \pm 0.011$ & $1.344 \pm 0.006$ & $87.9\%$  \\
\hline \multicolumn{5}{l}{\textbf{EA} } \\ \hline
\quad RT& $0.781 \pm 0.007$ & $0.797 \pm 0.007$ & $0.797 \pm 0.007$ & $*$  \\
\quad RR& $1.093 \pm 0.008$ & $0.995 \pm 0.020$ & $1.105 \pm 0.008$ & $112.4\%$  \\
\quad BA& $1.263 \pm 0.008$ & $1.140 \pm 0.051$ & $1.275 \pm 0.008$ & $109.8\%$  \\
\quad WS& $1.262 \pm 0.009$ & $1.042 \pm 0.013$ & $1.233 \pm 0.010$ & $86.6\%$  \\
\hline \multicolumn{5}{l}{\textbf{RFIM} } \\ \hline
\quad RT& $1.296 \pm 0.006$ & $1.299 \pm 0.006$ & $1.298 \pm 0.006$ & $*$  \\
\quad RR& $1.527 \pm 0.007$ & $1.530 \pm 0.007$ & $1.528 \pm 0.008$ & $*$  \\
\quad BA& $2.019 \pm 0.007$ & $2.020 \pm 0.007$ & $2.012 \pm 0.007$ & $*$  \\
\quad WS& $2.025 \pm 0.009$ & $1.990 \pm 0.008$ & $2.014 \pm 0.010$ & $67.3\%$  \\
\hline \end{tabular}}

	\caption{Results for four different problems: maximum cut (MaxCut), binary spin glass (BinSG), Edwards-Anderson (EA) and random field Ising model (RFIM).
	Each problem is studied on four different topologies: random recursive trees (RT), random $3$-regular graphs (RR), Barabasi-Albert graphs with $m=2$ (BA), and Watts-Strogatz graphs with $k=4, \ p=0.1$ (WS).
	All graphs have $n=8192$ nodes.
	We report the solution quality from run-time constrained simulated annealing (SA), the min-sum algorithm, and the min-sum-GNN.
    To estimate the quality of a method we average over $100$ problem instances and on each instance take the best of $10$ trials.
	The min-sum algorithm systematically performs worse than simulated annealing, except on trees where it is exact, or RFIM on locally tree-like graphs where it is asymptotically correct.
    The min-sum-GNN improves over min-sum in all cases where min-sum is not approximately correct, closing more than $80\%$ of the gap between min-sum and simulated annealing in all cases except the RFIM-WS. 
    Simulated annealing is run for a time roughly comparable to min-sum-GNN and the values are not saturated---running annealing for longer improves solution quality further.
	}
  \label{tab:results}
\end{table*}

The pattern of results is consistent across the first three problems---MaxCut, BinSG, EA.
On trees the min-sum algorithm is exact; no method can surpass it.
On other topologies simulated annealing outperforms min-sum, while the min-sum-GNN closes a considerable fraction of the gap.
The failure of the min-sum on locally tree-like graphs must be due to a failure to correctly account for the long cycles.
The min-sum-GNN appears to be doing something to correct for this.

The random field Ising model provides an interesting comparison study. With ferromagnetic bonds, finding the RFIM ground state is a minimum cut problem, which is computationally easy.
Min-sum might make mistakes but on locally tree-like graphs these are expected to be asymptotically small.
We find that min-sum is as good or better than annealing on locally tree-like graphs.
However, Watts-Strogatz graphs are not locally tree-like; the min-sum-GNN improvement over the min-sum is presumably a result of correcting for short cycles rather than the long ones.

Overall, the min-sum-GNN is an effective method. It finds solutions of considerably better quality than the GNNs that fail to beat simple methods.
At the same time the learned method still falls slightly short of simulated annealing and for this reason we are not yet convinced it is a reliable tool in practice.

\section{Discussion}
As noted in Refs.~\cite{boettcher_inability_2022, angelini_modern_2022}, naive applications of GNNs to combinatorial optimization do not appear to be useful.
However, carefully constructing a model can significantly improve performance.
Incorporating domain knowledge---ideas from the physics of random graphs---leads to relatively simple adjustments that considerably improve performance.

While GNNs are typically shallow, deep models are required for competitive optimization performance.
By incorporating the Laplacian mean-field approximation into the weight initialization, deep GNNs do not exhibit oversmoothing.
By mimicking the energetic cavity argument to improve the message passing architecture, we find GNNs that are much closer to simulated annealing in performance.
Good neural network construction builds atop traditional methods.

The min-sum-GNN improved over the base method and could match simulated annealing at similar run times for several tasks.
However, simulated annealing makes many fewer assumptions and one can easily improve performance by increasing the runtime.

Additionally, neural network approaches are ``learned'', i.e., they contain many free parameters that must be fit for a particular objective.
This process is a considerable upfront expense---one that we have not accounted for in our experiments.
Worse, even if we could amortize training costs we face out-of-distribution issues:
whenever an instance is substantially different to the training set, performance may be badly degraded.
Simulated annealing has no equivalent shortcoming; it can readily be applied to arbitrary problems and topologies. 

Learned optimization methods might be useful for finding solutions to a guaranteed fixed distribution of problems, under strict time constraint,
but this is a very narrow use case.
For now, we do not see a compelling reason to deploy GNNs for these combinatorial optimization tasks.
If and when that changes we suspect it will be due to a new architecture that successfully leverages heuristics.

What new physics can we learn from these neural networks? Our attention should also turn to inverting this process. It is well-established that the energetic cavity equations make mistakes due to the presence of cycles.
Our key finding is that there \emph{exists} a set of cavity equations, of the form of Eq.~\eqref{eq:minsum_GNN}, which considerably correct for shortcomings.
Characterizing these equations and explaining them from first principles should be the goal of future work.

\section*{Data Availability}
Code implementing our methods is freely available at \href{https://github.com/joebacchus/GNN-cavity-opt}{https://github.com/joebacchus/GNN-cavity-opt}.

\begin{acknowledgments}
GTC thanks Luana Ruiz for valuable conversations which helped to shape this project, and the hospitality of the Simons Institute for the Theory of Computing during the program \emph{Graph Limits and Processes on Networks: From Epidemics to Misinformation}.
JBG acknowledges funding from the Engineering and Physical Sciences Research Council through studentship 2926881.
\end{acknowledgments}

\appendix

\section{Experimental procedure} \label{app:training}
The GNNs are trained by generating new random instances of the given problem.
Instances are grouped into batches of size $2^{15}/n$. 
Training runs for $2000$ steps using the Adam optimizer \cite{kingma_adam_2015}.
The learning rate at step $t$ is set $\eta_t = 0.0003 \times 0.998^{t}$.
GNNs run at depth $T=256$ and width $d=32$, unless otherwise stated, and use the default initialization parameters of the PyTorch Geometric library \cite{fey_fast_2019}.

After training, to estimate the quality of a method we average over $100$ problem instances.
When a method is probabilistic then on each instance we use the best of $10$ trials.


\end{document}